\documentclass[onecolumn]{aa}
\usepackage{graphicx}

\begin{document}
\title{IBIS/PICsIT in-flight performances\thanks{Based on observations with
INTEGRAL, an ESA project with instruments and science data centre funded by ESA member
states (especially the PI countries: Denmark, France, Germany, Italy, Switzerland, Spain),
Czech Republic and Poland, and with the participation of Russia and the USA.}}

\author{G.~Di~Cocco, E.~Caroli, E.~Celesti, L.~Foschini, F.~Gianotti, C.~Labanti, G.~Malaguti, A.~Mauri, E.~Rossi, F.~Schiavone, A.~Spizzichino, J.B.~Stephen, A.~Traci, M.~Trifoglio}

\offprints{G. Di Cocco  - \email{dicocco@bo.iasf.cnr.it}
}

\institute{Istituto di Astrofisica Spaziale e Fisica Cosmica (IASF) del CNR - Sezione di Bologna\\
Via Gobetti 101, 40129 Bologna (Italy)}

   \date{Received 15 July 2003; accepted 2 September 2003}

   \abstract{PICsIT (Pixellated Imaging CaeSium Iodide Telescope) is the high energy detector of the IBIS telescope on--board the INTEGRAL satellite. PICsIT operates in the gamma--ray energy range between $175$ keV and $10$ MeV, with a typical energy resolution of $10$\% at $1$ MeV, and an angular resolution of 12 arcmin within a $\sim$100 square degree field of view, with the possibility to locate intense point sources in the MeV region at the few arcmin level. PICsIT is based upon a modular array of 4096 independent CsI(Tl) pixels, $\sim$0.70 cm$^2$ in cross--section and 3 cm thick. In this work, the PICsIT on--board data handling and science operative modes are described. This work presents the in--flight performances in terms of background count spectra, sensitivity limit, and imaging capabilities.
   }
\authorrunning{Di Cocco et al.}
\titlerunning{IBIS/PICsIT in--flight performances}

   \maketitle
%
%________________________________________________________________

\section{Introduction}
The IBIS telescope (Ubertini et al. 2003) provides unprecedented imaging capabilities over three decades of energy intervals in the gamma--ray regime, using a composite two--layer detection system made of discrete independent pixels. The upper layer, ISGRI (IBIS Soft Gamma--Ray Imager), is the low energy detector (Lebrun et al. 2003). ISGRI is based on CdTe pixels and covers the energy band $15-1000$ keV. Placed 10 cm below ISGRI, PICsIT (Pixellated Imaging CsI Telescope), is based on CsI(Tl) crystals, and it is operative between $175$ keV and $10$ MeV. Both layers are position sensitive and allow the reconstruction of the shadowgram cast by the tungsten coded mask placed $320$ cm above ISGRI. At that distance, one mask pixel, 1.12 cm wide, subtends an angle of $\simeq$12 arcmin, which defines the angular resolution of the instrument, while the point source location accuracy (PSLA) for strong sources ($\sim20\sigma$) is about 1 arcmin. A passive collimation system is present between ISGRI and the coded mask to shield the detectors against the cosmic diffuse background up to about 300 keV, while both layers are in anticoincidence with an active BGO veto system which surronds the sides and bottom of PICsIT for high--energy background rejection (Quadrini et al. 2003). The two--layer focal detection plane array is also coupled with the on--board IBIS calibration unit based on a $^{22}$Na source with a modified veto element as tagging detector, for gain and performance monitoring (Bird et al. 2003a).

PICsIT (Labanti et al. 2003) has been designed to attain, with the same instrument, good sensitivity, fine imaging, and moderate energy resolution in the gamma--ray energy domain between 175 keV and 10 MeV. It consists of 8 totally independent rectangular modules arranged in a $2\times4$ array inside an aluminium mainframe. Each module contains 512 ($16\times32$) CsI(Tl) scintillation crystals, each one coupled with a silicon photodiode (PD) for fluorescence light collection. The total assembly is a 4096 ($64\times64$) pixel matrix. Each pixel is $0.84\times0.84$ cm$^2$ in cross--section, and 3 cm thick. Every crystal is polished and wrapped with a white diffusive paper to provide stability and physical isolation between pixels. Each PD is interfaced to an analogue front--end electronic chain including a biasing filter, charge amplifier, shaping amplifier, threshold discriminator, and peak detector plus stretcher. These circuits are implemented in an analogue application specific integrate circuit (ASIC). Each ASIC receives the signals from 16 PDs. The key performance parameter of PICsIT are listed in Table 1.

The PICsIT detector has undergone a long series of functional tests and scientific calibration before and after the the integration in the IBIS telescope. The pixel--by--pixel channel--to--energy relationship was calibrated on--ground in order to create the look--up tables (LUT) uploaded in the on--board data handling system (Labanti et al. 2003). The LUTs have then been re--created using the in--flight data from the IBIS calibration unit (Bird et al. 2003), and the first in--flight calibration results indicate a very good stability of the behaviour of PICsIT (Malaguti et al. 2003).

\begin{table}
\begin{center}
\label{perf}
\caption{IBIS/PICsIT key performance parameters.}
\vspace{0.1cm}
\begin{tabular}{l c l }
\hline\hline
Parameter             & Value               & Notes                        \\
\hline\hline
Detection Area        & 2890 cm$^2$         & $4096\times0.84\times0.84$   \\ \hline
Effective Area        & $\sim$1400 cm$^2$         & at 500 keV                   \\
                      & $\sim$ 600 cm$^2$         & at 2000 keV                  \\ \hline
Energy Range          & 175 keV -- 6.5 MeV   & single events                \\
                      & 350 keV -- 13 MeV    & multiple events              \\    \hline
Energy Resolution     & 18\% at 511 keV     & in--flight results   \\
(corrected for offset)& 9\% at 1275 keV     &                             \\ \hline
Background Rate       & 2900 cts/s          & single events                \\
(Revolution 38)       & 650 cts/s           & multiple events              \\ \hline
Angular Resolution    & $12'$               & sampled in $10'$ pixels     \\ \hline
PSLA                  & 5$'$                & {}         \\ \hline
Continuum Sensitivity & $\sim2\times10^{-6}$  photons cm$^{-2}$ s$^{-1}$ keV$^{-1}$ & at 500 keV                   \\
(3$\sigma$, 10$^5$ s, $\Delta$E=E/2)
                                    & $\sim1\times10^{-6}$  photons cm$^{-2}$ s$^{-1}$ keV$^{-1}$ & at 2 MeV                 \\ \hline
Line Sensitivity      & $\sim2.5\times10^{-4}$ photons cm$^{-2}$ s$^{-1}$ & at 500 keV                   \\
(3$\sigma$, 10$^5$ s, $\Delta$E=E/2) & $\sim2.6\times10^{-3}$ photons cm$^{-2}$ s$^{-1}$  & at 2 MeV                 \\
\hline\hline
\end{tabular}
\end{center}
\end{table}

\section{IBIS/PICsIT modes of operation and in--flight context}
The on-board telemetry limitations do not allow a complete (position, time, energy), event--by--event, transmission of PICsIT data.
This has implied a significant amount of on--board processing.
Each PICsIT valid signal\footnote{Signals are not treated as PICsIT events when they are in coincidence with ISGRI ones (Compton events), calibration unit tags (calibration events), or veto strobes (removed as background).} is first corrected for gain non--uniformity by means of the on--board LUT. These corrected events are then associated with their respective energy deposit, incidence pixel, and time tag.
This on--board event pre--processing is performed differently according to the number of pixels triggered, the selected PICsIT operative science mode, and depends finally on the status of an on--board set of parameters which can be updated and uploaded via telecommand.

\subsection{Event multiplicity and handling}
In order to minimize the impact on scientific performances caused by the strong limitation imposed by the telemetry budget at satellite level, PICsIT data are treated on--board in different ways depending upon their {\em multiplicity}. A {\em single event} is the result of a gamma--ray photon triggering only one pixel, while a {\em multiple event} triggers two ({\em double}) or three ({\em triple}) pixels. Events with multiplicity greater than 3 are not transmitted to ground. Singles and multiples are put in two independent data sets. Multiple events are then processed on--board by the data handling system (DHS) and associated with one single energy deposit and incidence pixel. The first is the sum of the energies deposited in each pixel, while the incidence pixel, to form the multiple event shadowgram to be coupled with the mask pattern for image deconvolution, is reconstructed on--board by the DHS, which calculates the most probable incidence position implementing an algorithm developed from MonteCarlo simulations and scientific calibrations (Malaguti et al. 2001).
The on--board event handling introduces a certain degree of detector non--uniformity. This is due to the fact that multiple events are treated at semi--module level ($16\times16$ pixels), and, since there is no electronic communication between semi--modules, energy deposits falling across semi--module boundaries are not recombined into a single event. This gives rise to characteristic bi--convex (concave) shapes of the background for multiple (single) events (Malaguti et al. 2001, Bird et al. 2003).

\subsection{PICsIT operative scientific modes}
PICsIT has two fundamental operative modes: {\em Photon--Photon Mode} (PPM) and {\em Standard Mode} (SM). Because of the already mentioned satellite telemetry limitations, PPM can be used only in the case of limited count rate (e.g. selecting only one region of the operative spectral range).
In PPM, the events (singles and multiples) are transmitted with full information: energy (1024 channels), pixel position, time tag between events. During slews, PICsIT is by default, switched to PPM.

SM is the nominal operating mode for PICsIT and it is composed of two submodes: {\em Spectral Imaging} (SI) and {\em Spectral Timing} (ST). In SI the events (singles and multiples) are integrated on--board for a characteristic time which is coupled to the duration of one pointing, i.e. $\sim$2000 s, to produce 3--dimensional histograms. SI histograms are optimized for spatial resolution. The raw 1024 channel spectra are rebinned to 256 channels, and one energy spectrum is obtained for every pixel. The final result is a $256\times64\times64$ histogram. The spectral re--binning is performed by means of a non--linear binning table (contained in the DHS context) in order to optimize the energy sampling.

On the other hand, ST is optimized for temporal resolution at the expense of the spatial information. PICsIT and the
on--board DHS have been designed to have up to 8 programmable energy bands, and down to 0.97 ms time resolution. The implementation of ST in full resolution would however imply, because of telemetry limitation, significant loss of data for IBIS as a whole.
The default (from end of June 2003) functioning foresees four energy bands ($\sim150\div200$ keV, $\sim200\div250$ keV, $\sim250\div350$ keV, $\sim350\div700$ keV) with 1.9 ms time resolution.

\subsection{PICsIT and DHS in--flight context}
PICsIT in--flight data processing requires a number of parameters and algorithm to be present and implemented on--board. The main ones are: LUT for gain and offset correction, channel--to--energy conversion, binning tables for SM, incidence pixel reconstruction algorithm from multiple events, dead pixel map. These are all stored in the PICsIT and DHS {\em contexts}. The contexts can be downloaded as part of the scientific housekeeping data (HKD) and their contents can be updated and uploaded via telecommand following a ``reboot'' of PICsIT.

A typical example is the handling of noisy pixels. When a pixel becomes noisy, it is killed by the on--board automatic noisy pixel handling system. If it remains noisy, its coordinates are put in the killed pixel map contained in the context and the pixel will remain switched off at the first PICsIT reboot. Only 2 pixels have become noisy during the first 7 months of the mission, and have therefore been killed, for a total of 51 killed pixels ($\sim$1\% of total detecting area).

\section{In--flight performance and sensitivity}
The level and stability of PICsIT scientific performance has been monitored during the first months of the mission using the data from the in--flight calibration unit. For the first 4 months of the mission, PICsIT was switched off during belt transits. This implied a temperature decrease of $\sim$10$^\circ$C at constant satellite attitude during belt exits. These temperature fluctuations did not have measurable influence on
energy resolution (measured at the two $^{22}$Na lines: 511 and 1275 keV), which remained constant. On the other hand, the observed gain drift during temperature increase and stabilization at belt exits up to revolution 37 are fully explained by temperature fluctuations (Malaguti et al. 2003). Therefore, after having verified the absence of significant effect on instrument behaviour, leaving few modules on during belt transits for several weeks, it was decided, starting from revolution 38, to keep PICsIT always switched on to avoid unnecessary temperature fluctuations and increase the instrument stability. The observed residual gain fluctuations are to be ascribed to the changing aspect of the satellite with respect to the sun direction, for the various pointings.

\subsection{PICsIT background count spectrum}
\label{sect_bkg}

%
%______________________________________________ Bkg
   \begin{figure}
   \centering
  \includegraphics[scale=0.4]{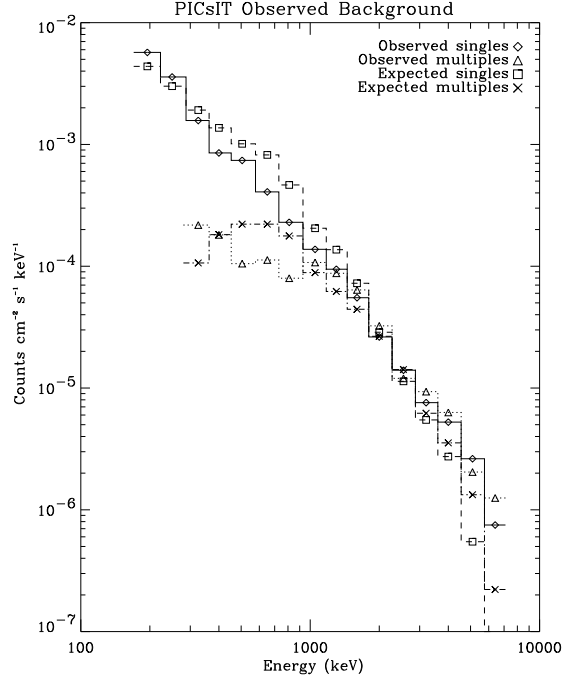}
   \vspace*{0.1cm}
   \caption{PICsIT in--orbit background spectrum for single and multiple events compared with the expected expected spectrum obtained from MonteCarlo simulations normalized to 7500 counts/s.}
    \label{PICsIT_bkg}
    \end{figure}

Prior to IBIS/PICsIT, no other pixellated CsI detector of this size was flown on a satellite. For this reason, the in--orbit background level and energy spectrum predictions, based on MonteCarlo simulations, or on semi--empirical scaling from other high energy space telescopes (e.g. OSSE, SIGMA), were accurate only within a factor of $\sim$2.

A series of photon-particle transport simulation programs have been developed to evaluate the in-orbit background rate and spectral shape. These resulted in  a predicted total (all events) count rate between 1500 and 2000 counts/s (Armstrong et al. 1999, Lei et al. 1997). However, PICsIT design assumed a conservative estimate of 7500 counts/s at solar maximum level in the energy range 140 keV -- 10 MeV. The most important contigency included in this evaluation is the uncertainty in the incident cosmic ray proton flux (Ferguson et al. 2003).
Figure \ref{PICsIT_bkg} shows the PICsIT in--orbit background spectrum for single and multiple events compared with the spectral shape obtained from Monte--Carlo simulations and scaled up to 7500 counts/s. The measurements are from the empty field acquisitions performed during revolution 38 (just before the first Crab observation), and give a total counting rate of $\sim$2900 and $\sim$650 counts/s for single and multiple events, respectively.

It is worth noting that, while the total count rate is a factor of $\sim$2 less than the expected conservative value, the observed background flux up to $\sim$250--300 keV is higher than expected. This is partly due to the events generated by the high--energy tracks to be ascribed to the photon--particle shower generated by high--energy cosmic rays interacting in the satellite--telescope structure or within PICsIT itself (Segreto et al. 2003).

%____________________________________Continuum sens
   \begin{figure}
   \centering
   \includegraphics[scale=0.45]{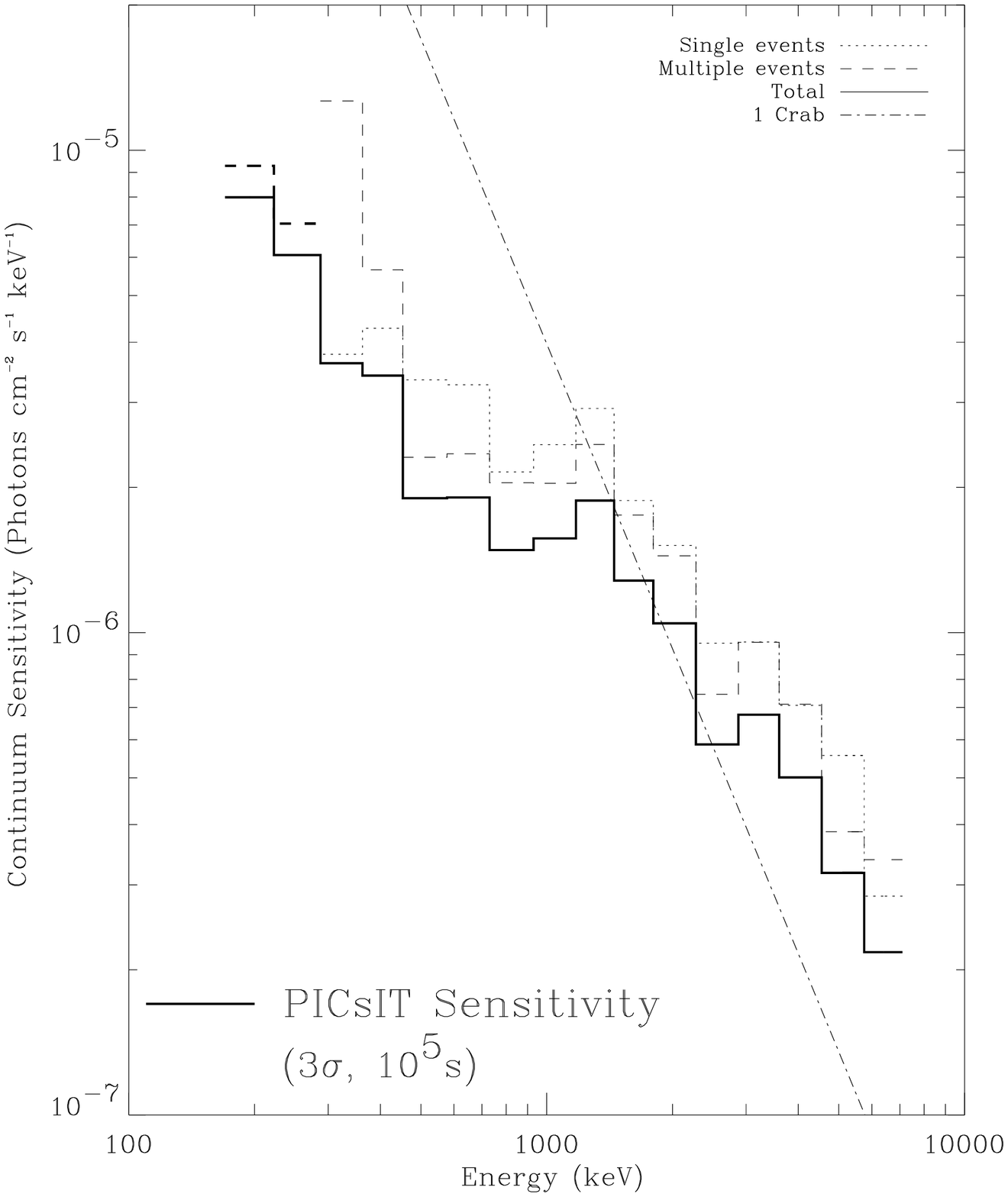}
   \includegraphics[scale=0.45]{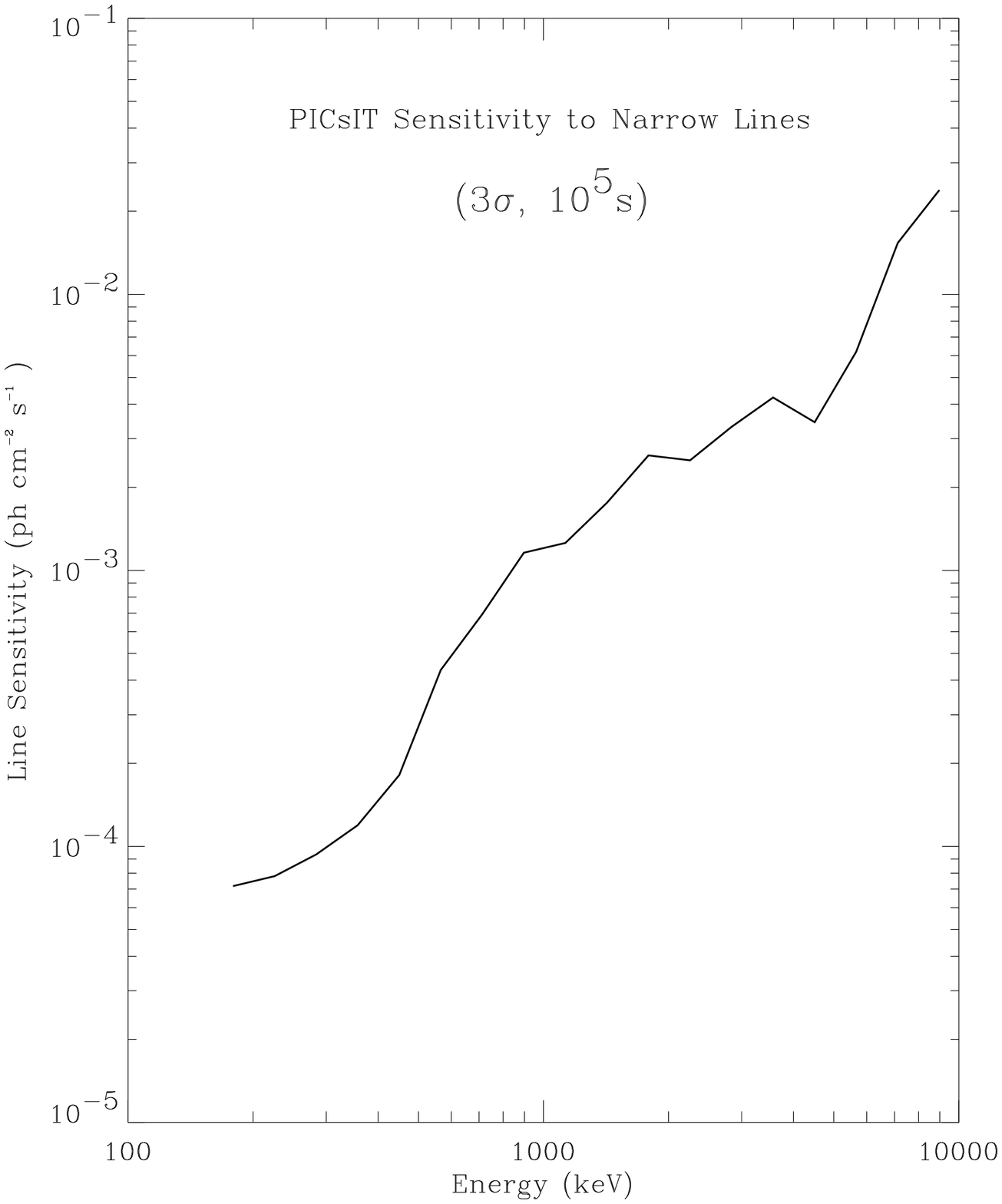}
   \vspace*{0.1cm}
   \caption{PICsIT minimum detectable flux for continuum (left panel) and narrow line (right panel) emission as a function of energy assuming only statistical effects, and the in--flight measured background. The dashed lines of the first two bins in the left panel indicate the worsening due to the effect of cosmic--ray induced tracks.}
    \label{Sens_Fig}
    \end{figure}

\begin{table}
\label{senstab}
\caption{IBIS/PICsIT parameters for the calculation of the statistical continuum flux sensitivity limit.}
\vspace{0.1cm}
\begin{tabular}{ c c c l }
\hline\hline
Parameter                  & Description              & Units                  & Notes                              \\
\hline\hline
$F_{\rm min}^{\rm Cont}$   & minimum detectable continuum flux  & photons cm$^{-2}$ s$^{-1}$ keV$^{-1}$   &                                    \\ \hline
$F_{\rm min}^{\rm Line}$   & minimum detectable line flux       & photons cm$^{-2}$ s$^{-1}$       &                                    \\ \hline
$n_{\sigma}$               & statistical significance           & -                      & 3                                  \\ \hline
$\varepsilon$              & total detection efficiency         & -                      & MonteCarlo and ground calibrations \\ \hline
$\varepsilon_P$            & photopeak detection efficiency     & -                      & MonteCarlo and ground calibrations \\ \hline
$T$                        & observing time                     & s                      & 10$^5$ s                           \\ \hline
$\Delta E$                 & Energy bin                         & keV                    & see Figure \ref{Sens_Fig}                 \\ \hline
$\delta E$                 & Energy resolution                  & keV                    & 19\% at 511 keV, 10\% at 1 MeV                                 \\ \hline
$\Theta_{\rm c}$           & Closed mask element transparency   & -                      & $\sim$6\% at 400 keV,
                                                                                           $\sim$31\% at 3.5 MeV  \\
\hline
$\Theta_{\rm o}$           & Open mask element transparency     & -                      & 1                                  \\ \hline
I                          & Coding power                       & -                    & 0.73$^{(a)}$                       \\ \hline
$\varepsilon_{\rm IPR}$    & Incident pixel reconstruction efficiency & -                       &                \\ \hline
B                          & Background count rate              & counts cm$^{-2}$ s$^{-1}$ keV$^{-1}$    & in--flight data                     \\
\hline\hline
\end{tabular}\\
$^{(a)}$In the case of position sensitive detectors coupled with coded masks, mask and detector have in general different spatial sampling, and the coding power $I$ is defined by: \( I=1-(L_{\rm D}/3L_{\rm M}) \) where $L_{\rm D}$ and $L_{\rm M}$ are the detector and mask pixel dimensions respectively (Skinner 1995). For PICsIT we have $L_{\rm D}/L_{\rm M} = 11.2/9.2$, and therefore $I=0.73$.
\end{table}

\subsection{Continuum and line sensitivity}
The statistical limits, i.e. the minimum detectable fluxes in case of negligible systematic effects, of PICsIT sensitivity to continuum ($F_{\rm min}^{\rm Cont}$ in photons cm$^{-2}$ s$^{-1}$ keV$^{-1}$), and narrow\footnote{In this context an emission line is considered narrow, when its width is small compared with the energy resolution of the instrument at that energy.} line ($F_{\rm min}^{\rm Line}$ in photons cm$^{-2}$ s$^{-1}$) emission, are given by:

\begin{equation}
\label{senseq}
F_{\rm min}^{\rm Cont.} =  \frac{I \cdot n_{\sigma}}{\varepsilon \cdot A \cdot T \cdot \Delta E \cdot (\Theta_{\rm o}-\Theta_{\rm c})^2 \cdot \varepsilon_{\rm ipr}} \left(n_\sigma(\Theta_{\rm o}+\Theta_{\rm c})+\sqrt{n_{\sigma}^{2} \cdot (\Theta_{\rm o}+\Theta_{\rm c})^2+4\cdot(\Theta_{\rm o}+\Theta_{\rm c})^2 \cdot A \cdot T \cdot \Delta E \cdot B}\right)
\end{equation}

\begin{equation}
F_{\rm min}^{\rm Line}  =  \frac{I \cdot n_{\sigma}}{0.75 \cdot \varepsilon_P \cdot A \cdot T \cdot (\Theta_{\rm o}-\Theta_{\rm c})^2 \cdot \varepsilon_{\rm ipr}} \left(n_\sigma(\Theta_{\rm o}+\Theta_{\rm c})+\sqrt{n_{\sigma}^{2} \cdot (\Theta_{\rm o}+\Theta_{\rm c})^2+4\cdot(\Theta_{\rm o}+\Theta_{\rm c})^2 \cdot A \cdot T \cdot \delta E \cdot B}\right)
\end{equation}

Since, both in PPM and SM, PICsIT single and multiple events are contained in two different and independent data sets, the total statistical limit for the minimum detectable flux is given by:

\begin{equation}
 F_{\rm min}^{\rm Tot} = \frac{1}{\sqrt{ \frac{1}{(F^{\rm singles}_{\rm min})^2} + \frac{1}{(F^{\rm multiples}_{\rm min})^2}}}
\end{equation}

The input parameters for equations \ref{senseq} are described in Table \ref{senstab}. The background spectrum is shown in figure \ref{PICsIT_bkg} and discussed in section \ref{sect_bkg}, the transparency of the closed elements of the coded mask has been calculated from the cross--section of 16 mm of tungsten, while the photopeak and total detection efficiency for single and double events has been obtained by means of a MonteCarlo photon--particle transport code cross--checked with the IBIS on--ground calibration results (Bird et al. 2003b).

Figure \ref{Sens_Fig} shows the continuum and line PICsIT sensitivity curves, based on statistical principles only and calculated using the in--flight measured background spectrum.
An important caveat is that the actual sensitivity limit depends also on systematic effects, mainly due to non-uniformity of the
spatial response for both background and source events (Natalucci et al. 2003), to cosmic-ray induced events (Segreto et al. 2003), and
to background temperature variations (Stephen et al. 2003). The actual version of the INTEGRAL Data Analysis System
(IDAS\footnote{IDAS software is available from http://isdc.unige.ch/index.cgi?Soft+dowload . See also Goldwurm et al. (2003) for details.})
performs the background correction on the basis of an empty field observation (Goldwurm et al. 2003). A refined correction
of these effects will be possible with a background model obtained after an accurate study of the in-flight data.

\subsection{Imaging capabilities}
\begin{figure*}
   \centering
   \includegraphics[scale=0.24]{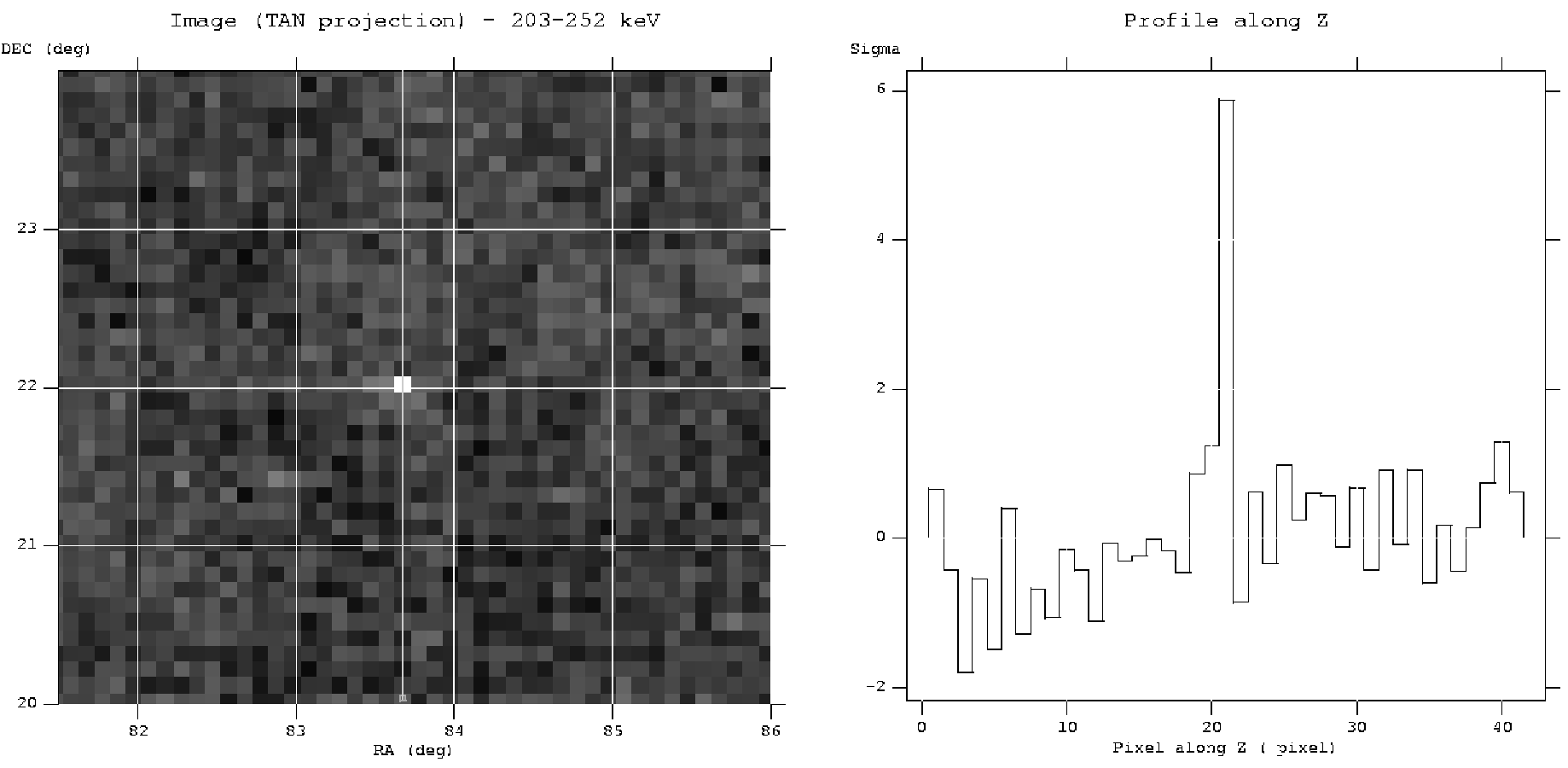}
   \includegraphics[scale=0.24]{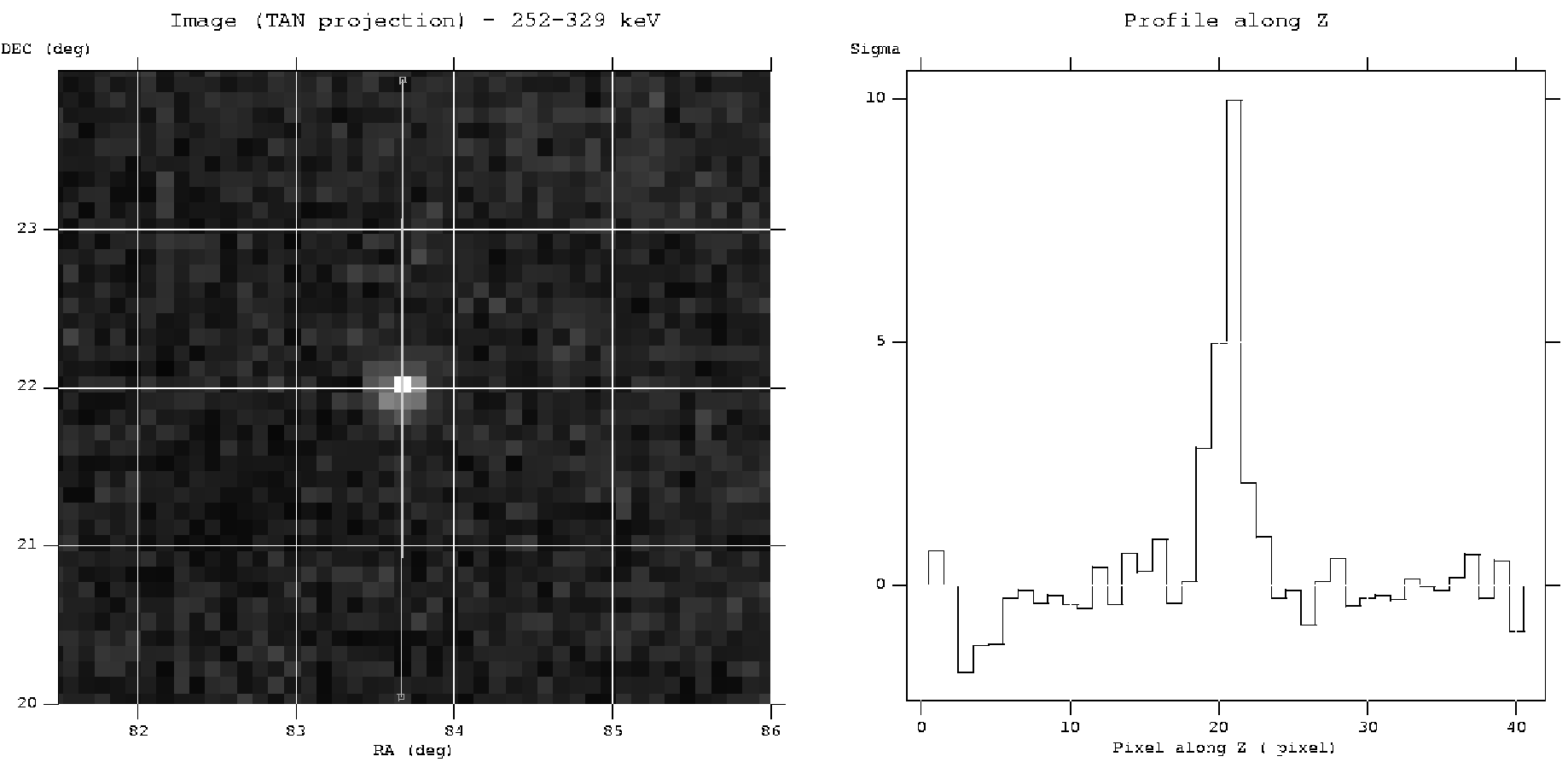} \\
   \includegraphics[scale=0.24]{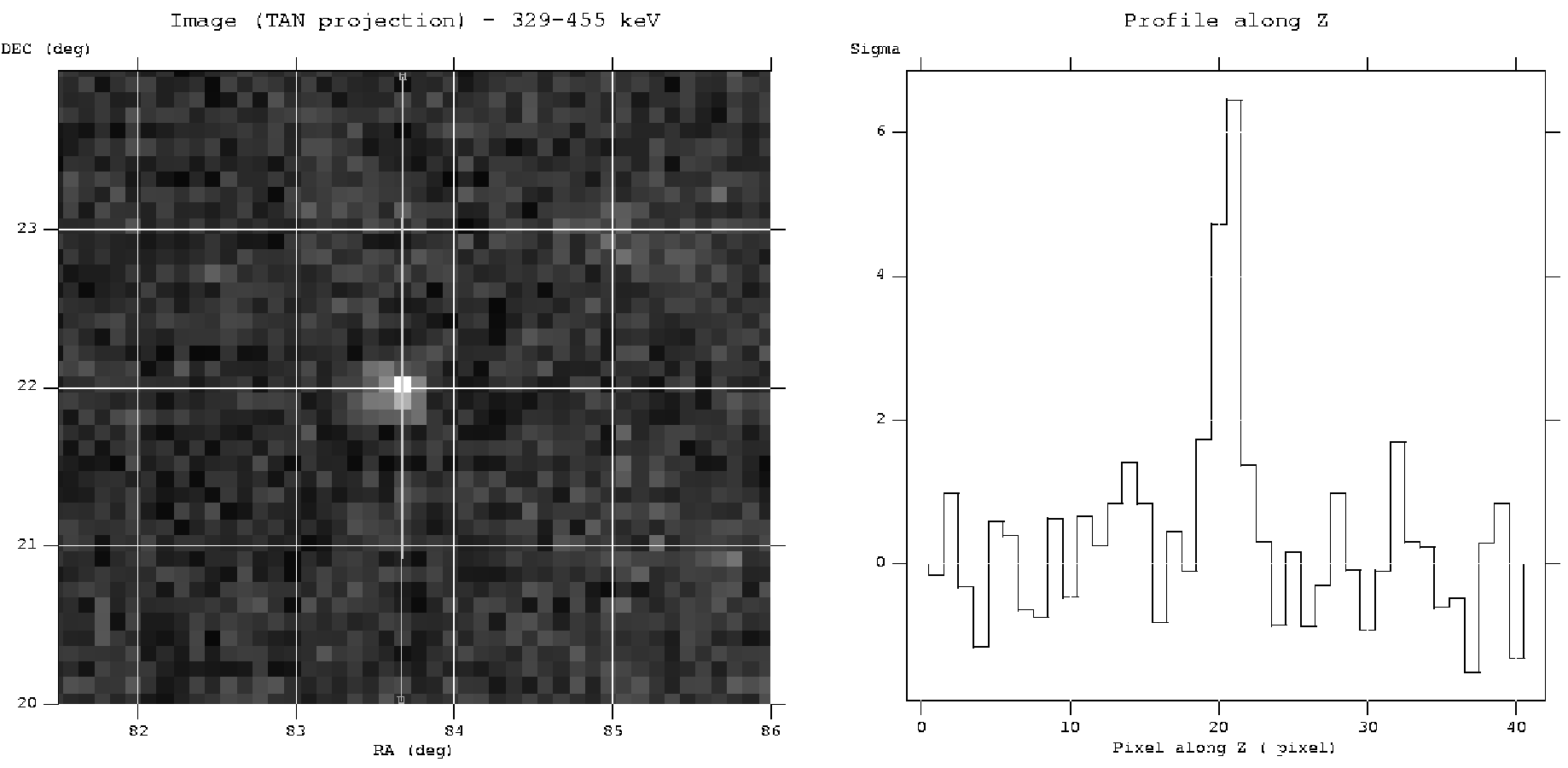}
   \includegraphics[scale=0.24]{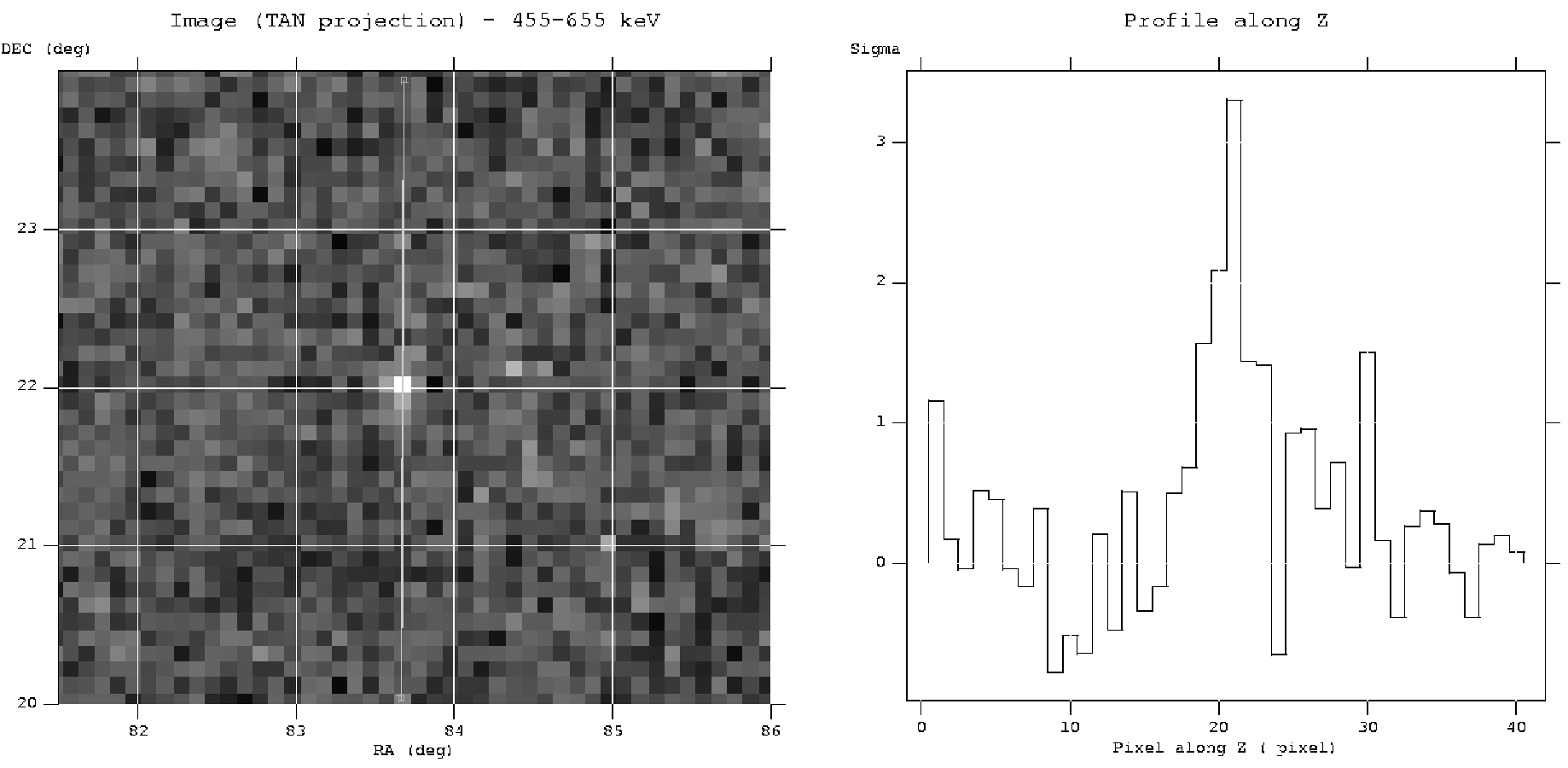}
   %\vspace*{0.5cm}
   \caption{Crab significance images and profiles obtained from observations performed during revolution 43 with the standard ISDC/ISSW. Deconvolved images are shown for 4 energy bands: $203-252$ keV, $252-329$ keV, $329-455$ keV, $455-655$ keV.}
    \label{CrabImages}
    \end{figure*}

\begin{figure*}
   \centering
   \includegraphics[scale=0.24]{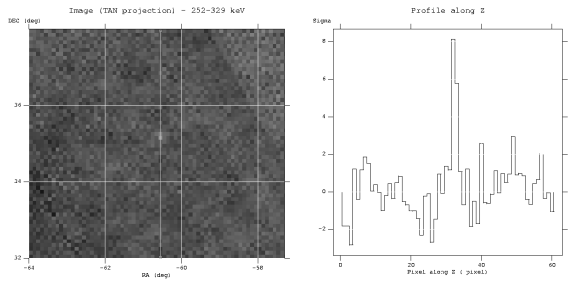}
   %\vspace*{0.5cm}
   \caption{Cyg X--1 significance image and profile in the $252-329$ keV band obtained from a 80 ks observation in revolution 11.}
    \label{CygX1}
    \end{figure*}

During the Performance Verification and Calibration Phases, IBIS has observed two point sources: Cyg X--1 and the Crab nebula.
The first results shown here have been obtained with the Offline Scientific Analysis (OSA) of the IDAS,
and show the imaging capabilities of PICsIT.
Figure \ref{CrabImages} and \ref{CygX1} show the significance map and profile in four energy bands
(only one for Cyg X--1), obtained decoding the shadowgram casted by the coded mask (see Goldwurm et al.
2003 for details regarding image deconvolution in IBIS and the definition of significance). The
significance of a detection in a coded mask telescope depends on a number of factors like, among others,
the ratio between detector plane spatial resolution and mask element size, background non--uniformities, and off--axis position of the source (Skinner 2002). In the current version of IDAS, the calculation of the significance takes into account statistical effects plus a first evaluation of background and detector non--uniformity (using the empty field observation of revolution 38), while an interpolation is implemented to account for the uneven sampling of mask elements by the detector pixels.

\begin{table}
\label{PSLA}
\caption{Actual source position accuracy. Columns: (1) Source name; (2) configuration during
the observation and offset with respect to the centre of the field of view; (3) Reference position
(RA, Dec, J2000) from catalog or circulars; (4) PICsIT position (RA, Dec, J2000);
(5) Offset [arcmin]. The position for PICsIT are taken in the $252-329$~keV energy band,
except for GRB021125 which has been obtained for $203-252$~keV, where the significance of the
detection was the highest.}
\vspace{0.1cm}
\centering
\begin{tabular}{lcccc}
\hline\hline
Source &  Configuration & Reference Position & PICsIT Position   & Offset \\
(1)    &  \multicolumn{1}{c}{(2)}           & (3)       & (4)      & (5)\\
\hline\hline
\multicolumn{5}{c}{Plain deconvolution (brightest pixel as reference)}\\
\hline
Crab       & staring, on--axis                   & $05:34:32;+22:00:52$ & $05:34:13;+21:55:48$ & $6.8'$\\
GRB 021125 & dithering, $7.3^{\circ}$ off--axis  & $19:47:56,+28:23:28$ & $19:47:34,+28:15:13$ & $9.5'$\\
\hline
\multicolumn{5}{c}{Calculation of the centroid}\\
\hline
Cygnus X--1 & dithering, on--axis                 & $19:58:22;+35:12:06$ & $19:58:08;+35:08:40$ & $4.4'$ \\
Crab        & dithering, on--axis                 & $05:34:32;+22:00:52$ & $05:34:24;+22:01:20$ & $1.9'$\\
Crab        & dithering, $30'$ off--axis          & {}                   & $05:34:17;+21:59:11$ & $3.8'$\\
GRB 021125  & dithering, $7.3^{\circ}$ off--axis  & $19:47:56,+28:23:28$ & $19:47:51,+28:19:16$ & $4.3'$\\
\hline\hline
\end{tabular}
\end{table}

Table 3 shows the reconstructed sky coordinates of the three point sources observed with PICsIT, compared with the reference position, which has been obtained from catalogues (Crab and Cyg X-1) and from ISGRI (GRB021125). Since PICsIT operates in a energy band where the signal-to-noise ratio (SNR) is intrinsically low, very long exposures are required. Therefore, it is not possible to make precise astrometry on a statistical basis. The residual offset can give an estimate of PICsIT point source location accuracy (PSLA), being the nominal value of the error radius of $5'$. A more detailed evaluation of PSLA (see Gros et al 2003 for more details) will be performed with the availability of a refined point spread function (PSF).

\begin{table}
\label{CrabFlux}
\caption{PICsIT observations of the Crab during Rev. 39 (staring, on axis, 77~ks),
Rev. 43 (dithering, on axis, 132~ks), Rev. 44 (dithering, off axis $0.5^{\circ}$, 104~ks).
The columns indicate the energy band [keV], the count rate in the peak pixel [counts/s]
and the significance of the detection for each revolution [$\sigma$].}
\vspace{0.1cm}
\centering
\begin{tabular}{ccccccc}
\hline
{}          & \multicolumn{2}{c}{Rev 39} & \multicolumn{2}{c}{Rev 43} & \multicolumn{2}{c}{Rev 44}\\
\hline
Energy Band & Rate & Significance  & Rate & Significance   & Rate & Significance  \\
\hline
$200-252$   & $2.5$  & $7.4$  & $1.5$ & $5.9$    & $3.1$ & $5.7$\\
$252-329$   & $2.2$  & $11.0$ & $2.1$ & $10.0$   & $2.6$ & $8.0$\\
$329-455$   & $1.3$  & $7.1$  & $1.2$ & $6.5$    & $1.4$ & $5.0$\\
$455-655$   & $0.7$  & $4.4$  & $0.6$ & $3.3$    & $0.6$ & $3.0$\\
\hline
\end{tabular}
\end{table}

Table 4 shows the reconstructed count rates and associated significance of the Crab observations in three different telescope configurations. In revolution 39, IBIS observed the Crab on--axis, staring on the source for 77 ks; in revolutions 43 and 44 the Crab was pointed with a $5\times5$ rectangular dither pattern (Jensen et al. 2003), on--axis for 132 ks and 30 arcmin off--axis for 104 ks, respectively.

The high fluctuation of the detected Crab rate in the lowest energy band is due to the presence of cosmic ray induced events. In fact, in SM the on-board accumulation of histograms allows only a partial {\em a posteriori} rejection of these events (see Goldwurm et al. 2003 for details on this type of correction).

Table 4 also shows that, for the 252--329 keV band, an increase of exposure does not result in the expected increase of detection significance (11$\sigma$ with 73 ks staring, against 10$\sigma$ with 132 ks dithering). This is explained by the fact that during revolution 43, the Crab was observed in dithering mode with the rectangular $5\times5$ pattern (Jensen et al. 2003). This pattern samples regions close to the edge of the fully coded field of view, thus implying a sensitivity degradation of $\sim$20\% (Foschini 2003).

These results are in good agreement with simulation results obtained by means of a dedicated MonteCarlo code in the energy range 252--655 keV, while are a factor of two lower than expected in the first energy channel, where the systematic effects caused by the cosmic--rays induced events affects the data analysis.

%
%______________________________________________ Crab
%

%
%______________________________________________ Cyg X-1

\subsection{Spectral \& Timing capabilities}
The spectral and timing characteristics of PICsIT were analysed
by using other sources. A solar flare, occurred on
November $10^{\mathrm{th}}$, 2002, at about $10:00$~UTC, allowing the
detection of several emission lines: the $e^{+}e^{-}$ annihilation
line at $511$~keV with the tail at lower energies made by the fusion
lines, the neutron capture at $2.2$~MeV, and the nuclear excitation lines
of $^{12}$C at $4.4$~MeV. These lines have confirmed the linearity of the
detector over the entire energy range, while a more detailed analysis will
be performed when the refined redistribution matrix (RMF) will be available.

The GRB of November $25^{\mathrm{th}}$, 2002, allowed a full test of the
imaging, spectral, and timing capabilities in photon--by--photon mode and
it is described in Malaguti et al. (2003).

In addition, the absolute timing performances both in standard and
photon--by--photon mode were analysed by using the Crab pulsar observations (see Kuiper et al. 2003).

\section{Conclusions}
The in--flight functional and scientific performances of the IBIS/PICsIT detector have been described and discussed,
together with the first observations performed during the PV and calibration phases.
The instrument has demonstrated to work nominally, from an engineering point of view with respect to the
design requirements. Activities are ongoing to improve the current scientific performances by minimizing
the systematic effects due to the background evaluation, cosmic--rays induced events, and detector
non--uniformities.

\begin{acknowledgements}
This work has been partially funded by the Italian Space Agency (ASI).
The italian participation to the INTEGRAL/IBIS project is financed by the
Italian Space Agency (ASI). Fruitful collaboration with LABEN S.p.A., Alenia Spazio,
and the use of the Calibration Facility at ESA/ESTEC is kindly acknowledged.
LF acknowledges the kind hospitality of the INTEGRAL Science Data Centre (ISDC)
during part of this work.
\end{acknowledgements}

\end{document}